\documentclass[twoside,english,10pt]{JAC2001}

\usepackage{amssymb}
\usepackage{graphicx}
\usepackage{babel}
\usepackage{multicol}

\begin{document}

\title{Vibrational Stability of NLC Linac accelerating structure}
\author{F.~Le~Pimpec, S.~Adiga (Stanford Univ.), F.~Asiri, G.~Bowden, E.~Doyle,
B.~McKee, A.~Seryi
\\SLAC, 2575 Sand Hill Road, Menlo Park CA 94025 , USA
\thanks{Work supported by the U.S. Department of Energy, Contract
  DE-AC03-76SF00515.}
\\S. Redaelli, ~~CERN, 1211 Geneva 23, Switzerland
\thanks{PhD student of the University of Lausanne,
Institut de Physique des Hautes Energies (IPHE), Switzerland.} }
\maketitle

\begin{abstract}
The vibration of components of the NLC linac, such as accelerating
structures and girders, is being studied both experimentally and
analytically. Various effects are being considered including
structural resonances and vibration caused by cooling water in the
accelerating structure. This paper reports the status of ongoing
work.
\end{abstract}


\section{Introduction}
As part of the R\&D effort for the Next Linear Collider (NLC), an
extensive program has been started to study the vibration induced by cooling water
on the NLC Linac and Final Focus components.

An adequate flow of cooling water to the accelerating structures is required in order to dissipate the heat load caused by the absorption of
RF power and to maintain the
structure at the designated operating temperature. This flow may cause vibration of the structure and its supporting girder. The acceptable tolerance for vibration of the structure itself is rather loose -- of the order of several micrometers.
The concern is that this vibration can couple to the linac quadrupoles either via the beampipe with its bellows or via the supports.
The vertical vibration tolerance for the linac quadrupoles is about 10~nm.

In this paper we focus on vibration of accelerating~RF structure
and girder induced by cooling water. Further papers will report on
vibration coupling to quadrupoles \cite{lepimpec:FNAL02} and on
investigation of additional girder damping
\cite{lepimpec:LINAC02}.

\section{Experimental setup}
\label{Expesetup}

For the vibration studies presented, an NLC design
accelerating structure \cite{zdr} was used. This structure is 1.8~m long and is
supported by a ``strongback'' of the same length. In the design it
was assumed that 3 such structures would be mounted on a single
6~m long girder. The required water flow is about $\sim$1~$\ell$/s
for each structure. It should be noted that the NLC currently
plans to use shorter RF structures than the one studied \cite{chris}.

\begin{figure}[tbph]
\begin{center}
\vspace{-0.3cm}
\includegraphics[clip=,width=7.3cm,totalheight=7cm]{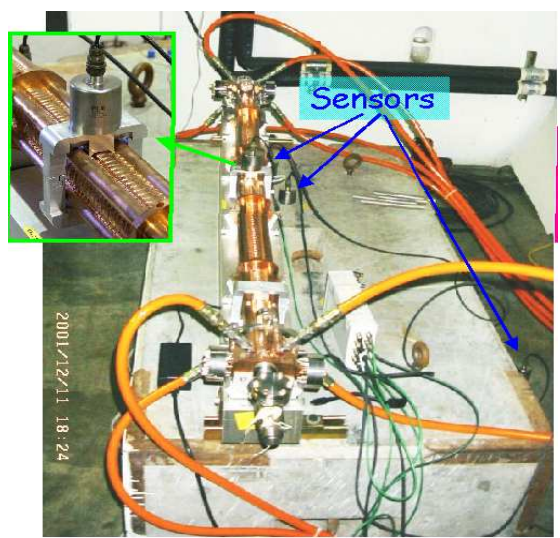}
\end{center}
\vspace{-0.7cm} \caption{RF structure and girder vibration setup
in the NLCTA area. } \vspace{-0.3cm} \label{NLCTAvibsetup}
\end{figure}

In the first set of experiments, we measured the vibration induced
by different flow rates passing through the structure-girder
system, as shown in Figure~\ref{NLCTAvibsetup}. The water was
supplied from the NLC Test Accelerator (NLCTA) area. The 1.8~m
long RF structure ($\sim$100 Kg) was mounted on a hollow aluminum
girder connected to a concrete block of $\sim$2225 kg. The block
was installed on rubber balls ($\sim$14Hz resonance) to isolate it
from the noisy floor of NLCTA. Vibration was monitored by four
piezo-accelerometers and one piezo-transducer was used to measure
water pressure fluctuations. The diameter of each of the four
cooling pipes were 1.9~cm. The flow of water and the pressure were
measured by a venturi tube and by two manometers installed at the
input and output supply.

The second set of experiments were designed to study the vibration caused
only by internal turbulence. The structure-girder was installed
in a much quieter place on the floor of the SLD collider hall and the water was gravity-fed from a tank located $\sim$18~m higher. The structure-girder was bolted to a $\sim$26T concrete block placed on a rubber mat ($\sim$30Hz
resonance). The maximum water flow through the structure was
limited to about 1.1~$\ell$/s.

In each experiment, it was possible to feed the water to the structure
either by using the 4 tubes separately or by feeding only 2
tubes on one end which were then connected to the adjacent tubes at the other end.
In the latter case, the flow in 2 tubes was in the opposite direction. In all the tests, only vertical vibrations were studied.

\section{Results and Discussion}
\label{Resultat}

\begin{figure}[tbp]
\centering
\includegraphics[clip=,width=7.5cm,totalheight=5cm]{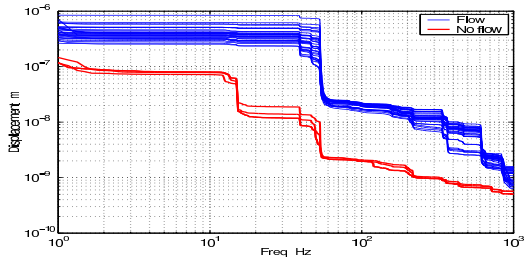}
\vspace{-0.4cm} \caption{Integrated displacement spectrum of the
structure with different flow rates(blue, top curves) and with no
flow (red, bottom curves). Several measurements are shown.} \vspace{-0.2cm}
\label{NLCTAvib01}
\end{figure}

\begin{figure}[tbp]
\centering \vspace{-0.2cm}
\includegraphics[clip=,width=0.85\columnwidth,totalheight=5cm]{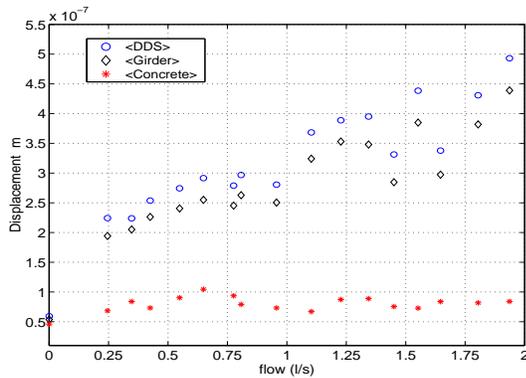}
\vspace{-0.4cm} \caption{Average integrated displacement above
10Hz of the structure (circles), of the girder (diamonds), and of
the concrete block(crosses).} \label{NLCTAvib02} \vspace{-0.5cm}
\end{figure}

The first results were obtained when the
structure was supplied by water from the NLCTA supply system
(Fig.\ref{NLCTAvibsetup}). Fig.\ref{NLCTAvib01} shows the
variation of the integrated displacement spectrum when the
structure is fed with different water flows, including no flow. In
the case without flow, both supply and return valves were closed.
In the case with flow, the supply valve was partially closed to
regulate the flow. Fig.\ref{NLCTAvib02} shows the average value
of the integrated displacement above 10Hz, for
different water flow rates.

\begin{figure}[t]
\begin{center}
\vspace{-0.0cm}
\includegraphics[clip=,width=7.5cm,totalheight=5cm]{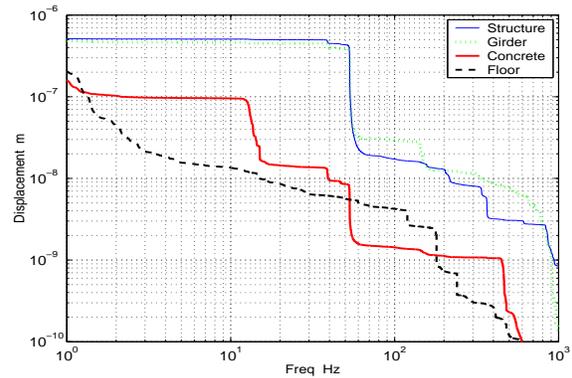}
\vspace{-0.8cm}
\end{center}
\caption{Integrated displacement spectrum of the structure,
girder, floor and concrete block with water flow of
$\sim$1.9~$\ell$/s in the NLCTA setup for a single measurement.}
\vspace{-0.5cm} \label{VibmaxflowNLCTA}
\end{figure}

Fig.\ref{VibmaxflowNLCTA} shows the integrated displacement of the
different elements of the system for a flow of $\sim$1.9~$\ell$/s
(twice nominal). One can see that the maximum amplitude of the
displacement comes from the peak at $\sim$~52Hz which is believed
to be a resonance of the hollow aluminium girder and RF structure
assembly. The 52Hz resonance is also seen in the motion of the
concrete block, due to coupling of the structure vibration to the
block.

\begin{figure}[tbph]
\begin{center}
\vspace{-0.3cm}
\includegraphics[clip=,width=7.5cm,totalheight=5cm]{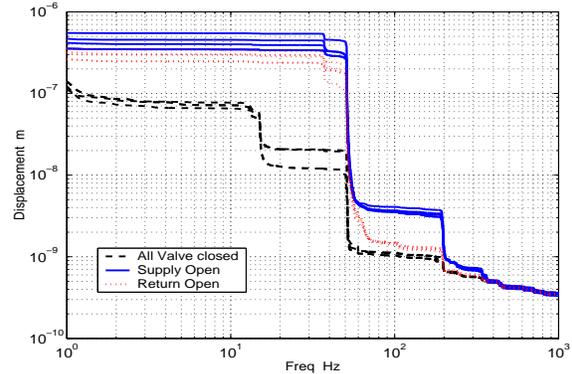}
\vspace{-0.8cm}
\end{center}
\caption{Integrated displacement spectrum of the structure with only
the supply open (blue solid line), only the return open (red
dotted line), and with all valves closed (black dashed line). In all cases, there is no water flow.} \vspace{-0.3cm} \label{NLCTAturbulence}
\end{figure}

The integrated displacement of the structure at maximum flow is
about 0.5~$\mu$m (see Fig.\ref{NLCTAvib02} and
\ref{VibmaxflowNLCTA}), while it is less than 0.08~$\mu$m when no
water flows through the structure. The rms amplitude of vibration
depends only slowly on the flow rate -- a 10-fold change in the
flow increases the vibration by only about a factor of 2. Note
that the system considered is above the turbulence threshold
(Re$>$2000) if the flow $>$~0.1~$\ell$/s, i.e. practically in all
the observed range. Such a slow dependence is explained if one
assumes that in the NLCTA setup, the structure vibrations were due
mostly to turbulence in the water supply system and hence to
pressure fluctuations in the incoming water. This assumption is
confirmed by Fig.\ref{NLCTAturbulence} which shows that merely
opening either supply or return valve, with the other valve
closed, produced an integrated displacement of 0.4~$\mu$m, almost
equal to the maximum displacement observed.

\begin{figure}[h]
\begin{center}
\vspace{-0.0cm}
\includegraphics[clip=,width=7.5cm,totalheight=5cm]{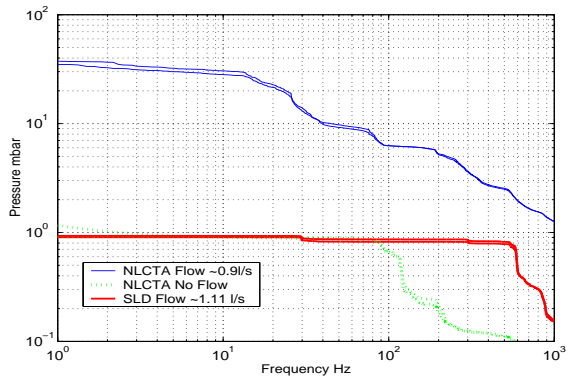}
\vspace{-0.7cm}
\end{center}
\caption{Integrated spectrum of incoming pressure fluctuations in
NLCTA water (top curves), in gravity fed water (middle curves),
and the sensor internal noise -- NLCTA case with all valves closed
(bottom curves).} \vspace{-0.4cm} \label{Presstransducer}
\end{figure}

Quantitatively, the incoming water can be characterized by the
spectrum of its pressure fluctuations. The integrated spectrum was
measured by a pressure piezo-transducer in the NLCTA setup as
shown in Fig.\ref{Presstransducer}. The incoming pressure causes
vibration of the RF structure through the force it exerts on the
unbalanced surface of the cooling pipes (usually equal to
cross-section of the pipe). It is interesting to note that the
spectrum of the incoming NLCTA water is rather smooth and does not
contain sharp peaks typically associated with the rotational
frequencies of pumps. This indicates that the turbulence itself,
and not the pumps, was the cause of the pressure fluctuations.

\begin{figure}[t]
\begin{center}
\vspace{0.1cm}
\includegraphics[clip=,width=7.5cm,totalheight=5cm]{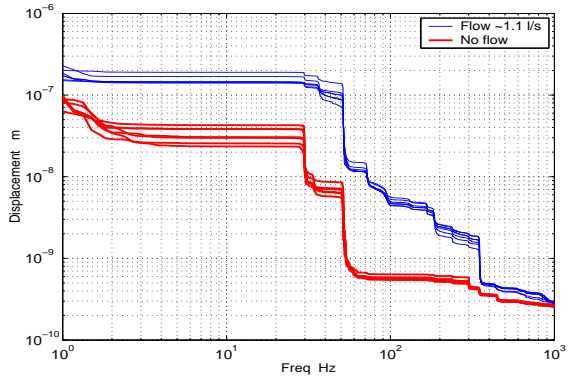}
\vspace{-0.7cm}
\end{center}
\caption{Integrated displacement spectrum of the structure
with the nominal water flow (blue) and with no flow (red)}
\vspace{-0.4cm} \label{SLDgravvib01}
\end{figure}

From the results obtained with a normal cooling water system, such
as the one used in NLCTA, it is clear that vibrations are caused
mostly by the turbulence of the incoming water, and not by
irrevocable internal turbulence in the accelerating structure
itself. To study the latter, the system was moved to the floor of
the SLD detector hall, and fed with ``quiet'' water coming from a
tank at about $\sim$18~m height. The maximum flow obtained in this
configuration was $\sim$1.1~$\ell$/s (equal to the nominal flow).

Figure~\ref{SLDgravvib01} shows the variation of the displacement
with and without flow obtained in the ``gravity fed'' experiment.
The RF structure vibration is $\sim$0.18~$\mu$m in this case, i.e.
about a factor of 2 smaller than in the NLCTA setup. The pressure
transducer shows that fluctuations in the gravity fed incoming
water are much smaller than in the NLCTA setup. It is believed
that this structure vibration was dominated by internal
turbulence.

\begin{figure}[t]
\begin{center}
{
\vspace{0.1cm}
\includegraphics[clip=,width=7.5cm,totalheight=5cm]{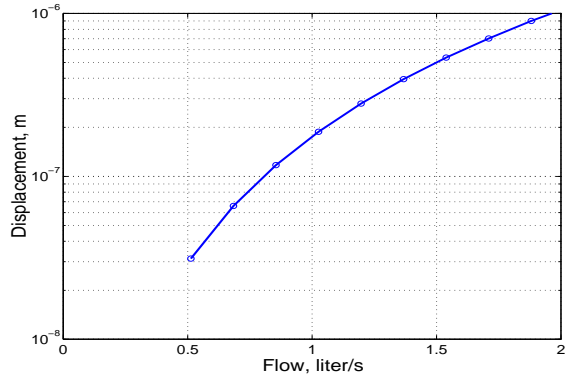}
\vspace{-0.7cm}}
\end{center}
\caption{Vibration of the RF structure at a resonance of 52~Hz
caused by internal turbulence in the cooling pipes as predicted by
a semi-analytical model \protect{\cite{Sri:2002}}.} \label{model}
\vspace{-0.4cm}
\end{figure}

While the dependence of vibration on flow in the gravity fed
experiment has not yet been measured, a semi-analytical model
\cite{Sri:2002} has been developed which predicts that the
amplitude will depend on the flow approximately as $flow^{2.5}$
(see Fig.\ref{model}). It is interesting to compare this
predictions with estimations from \cite{Schnell:2001kf} where the
amplitude of vibration caused by internal turbulence almost does
not depend on the flow (the dependence is only as $1/flow^{1/8}$).

The model used \cite{Sri:2002} is based on analytical or measured
mode-shape functions of structural resonances for the system
considered, as well as on empirical models of power spectrum and
spatial coherence properties of pressure fluctuations in turbulent
water \cite{auyang}. For nominal flow, the model predicts
$\sim$180~nm amplitude which is in good agreement with the gravity
fed experimental data, while for twice the nominal flow its
prediction is higher than even what was measured in NLCTA. It is
worth noting that most of parameters of the model (including
quality factor of the resonance) are difficult to predict
analytically. Therefore, caution is required in extending any such
model to a different system or parameter range.

Finally, in the gravity fed case, the vibration of a quadrupole
connected to the structure via a bellows was also measured and
found to be about 6~nm, which is tolerable for NLC. Ongoing work
on optimization of the design and on increasing structural damping
\cite{carter} is expected to further reduce the coupling of
vibration to the quadrupoles.

\section{Conclusion}

Cooling water can cause vibration of the accelerating structure both
through internal turbulence in the cooling
pipes in the structure, and through pressure fluctuations in the
supply water (external turbulence). The latter does not depend
on the flow rate through the structure and can be the dominant
source of vibration in practical situations. For the case studied, mechanical resonances of the
structure-girder assembly explain the measured amplitudes.
Vibration coupling to a quadrupole was found to be adequately
small. Optimization of design and increased mechanical damping is
expected to further reduce vibration.

\section{Acknowledgments}
We would like to thank Ralph Assmann, Marty Breidenbach, Domenico
Dell'Orco, Harry Carter, Tor Raubenheimer, Nikolai Solyak,
Cherrill Spencer, Nancy Yu, for help and useful discussions.


\begin{thebibliography}{9}

\bibitem{lepimpec:FNAL02}
{F. Le Pimpec, et al.}
\newblock In {\em {NLC US collaboration Workshop,}} FNAL, May, 2002,
\newblock {http://www-project.slac.stanford.edu/\\lc/local/Reviews/2002\_Spring-Collaboration/\\FNAL\_vibration\_may2002.pdf}.
\\[-5mm]

\bibitem{lepimpec:LINAC02}
{F. Le Pimpec et al.}
\newblock {submitted to LINAC 2002, Korea.}
\\[-5mm]

\bibitem{zdr}
{NLC ZDR Design Group}.
\newblock SLAC Report-474, 1996.
\\[-5mm]

\bibitem{chris}
{C. Adolphsen, in these Proceedings.}
\\[-5mm]

\bibitem{Sri:2002}
{S.~Adiga}.
\newblock {\em {Estimations of the cooling water induced vibrations in pipes and
  accelerating structures}}.
\newblock Technical report, SLAC To be published, 2002.
\\[-5mm]

\bibitem{Schnell:2001kf}
{W.~Schnell}.
\newblock  CERN-CLIC-NOTE-468, 2001; \\
also S.~Redaelli et al., in these Proceedings.
\\[-5mm]

\bibitem{auyang}
{M.K. Au-Yang}.
\newblock {\em {Flow-induced vibration of power and process plant components}}.
\newblock {ASME Press, New York}, 2001.
\\[-5mm]

\bibitem{carter}
{H. Carter, et al., FNAL, ongoing work.}
\\[-5mm]

\end{thebibliography}
\end{document}